# A Comparison of Mechanisms for Integrating Handwritten and Generated Code for Object-Oriented Programming Languages

Timo Greifenberg[1], Katrin Hölldobler[1]*, Carsten Kolassa[1], Markus Look[1], Pedram Mir Seyed Nazari[1], Klaus Müller[1], Antonio Navarro Perez[1], Dimitri Plotnikov[1], Dirk Reiss[2], Alexander Roth[1], Bernhard Rumpe[1], Martin Schindler[1] and Andreas Wortmann[1]

[1] *Software Engineering, RWTH Aachen University, Germany*

[2] *Institute for Building Services and Energy Design, TU Braunschweig, Germany*

{greifenberg, hoelldobler, kolassa, look, nazari, mueller, perez, plotnikov, roth, rumpe, schindler, wortmann}@se-rwth.de, d.reiss@tu-bs.de

Abstract: Code generation from models is a core activity in model-driven development (MDD). For complex systems it is usually impossible to generate the entire software system from models alone. Thus, MDD requires mechanisms for integrating generated and handwritten code. Applying such mechanisms without considering their effects can cause issues in projects with many model and code artifacts, where a sound integration for generated and handwritten code is necessary.
We provide an overview of mechanisms for integrating generated and handwritten code for object-oriented languages. In addition to that, we define and apply criteria to compare these mechanisms. The results are intended to help MDD tool developers in choosing an appropriate integration mechanism.

## 1 INTRODUCTION

Model-driven development (MDD) (Kleppe et al., 2003) pursues the vision to create complex software systems from abstract models by transforming these into concrete implementations (France and Rumpe, 2007). However, the prevailing conjecture is that deriving a non-trivial, complete implementation from models alone is not feasible (Wile, 2003). Current MDD techniques thus require tool developers to integrate generated and handwritten code. To perform this code integration, various mechanisms can be used. However, there is no best integration mechanism which should always be used. Instead, it depends on the context in which this code integration has to be carried out and on the concrete requirements which integration mechanisms are best suited to be applied.

To support MDD tool developers in selecting integration mechanisms, we examined existing mechanisms to integrate generated and handwritten code for object-oriented programming (OOP) languages. Additionally, we created a set of criteria focusing

*K. Hölldobler is supported by the DFG GK/1298 Algo-Syn.

on different properties of code integration mechanisms to assess and compare these mechanisms. The presented criteria are based on a decade of experiences in object-oriented software engineering and MDD research (Rumpe, 2011; Rumpe, 2012), code generator development and code integration research (Rumpe et al., 2010; Schindler, 2012), and experiences with MDD processes within various domains including automotive (Grönniger et al., 2008), cloud computing (Navarro Pérez and Rumpe, 2013), robotics (Ringert et al., 2013), and smart buildings (Kurpick et al., 2012).

We introduce eight handwritten code integration mechanisms and evaluate each with respect to our criteria. Moreover, we show strengths and weaknesses of each integration mechanism in the evaluation results. By means of this, we seek to increase the comparability between the integration mechanisms. In particular, this overview is intended to be used by MDD tool developers to find a proper integration mechanism on a case-by-case basis.

Please note, that the list of integration mechanisms and evaluation criteria presented in this paper does not claim to be complete. However, if further integration mechanisms need to be compared or the mechanisms need to be evaluated with respect to additional evaluation criteria, this paper can be used as a basis easily.

In summary, the contributions of this paper are:

- A list of evaluation criteria for code integration mechanisms.
- A collection of mechanisms to integrate generated and handwritten code.
- An evaluation of the integration mechanisms based on the list of evaluation criteria.

In the remainder, we introduce criteria to assess the different integration approaches (Section 2). The code integration approaches are separated into mechanisms based on specific concepts in programming languages (Section 3) and mechanisms free of such requirements (Section 4). Subsequently, we summarize and discuss the evaluation results (Section 5). After that we elaborate on related work (Section 6) before we conclude this contribution (Section 7).

## 2 EVALUATION CRITERIA

The following criteria concern different properties of integration mechanisms and aim at helping developers in selecting a suitable mechanism. These criteria are based on existing literature (Stahl and Völter, 2006; Pietrek et al., 2007), best practices in software development (Parnas, 1972; Gamma et al., 1995), and our experience (Rumpe, 2011; Grönniger et al., 2008; Rumpe et al., 2010; Kurpick et al., 2012; Rumpe, 2012; Schindler, 2012; Navarro Pérez and Rumpe, 2013; Ringert et al., 2013). This list does not claim to cover all aspects of handwritten code integration. Nonetheless, it can be used as an initial list of criteria which can be adapted to fit personal needs. The presented criteria are not weighted on purpose since a weighting is highly subjective and also tailored to an application scenario, which is not intended.

**C1: Separation of generated and handwritten code.** - Can generated and handwritten code be separated into different files?

Separation of concerns is an essential design practice in software development (Parnas, 1972) and has been proposed as a criterion to evaluate integration mechanisms for handwritten code by (Stahl and Völter, 2006; Pietrek et al., 2007). One crucial benefit of separating generated and handwritten code into different files is that it can be ensured that the generator does not overwrite handwritten code. In case of mixing generated and handwritten code in one file, the handwritten code might not always be preserved.

**C2: Support for overriding generated parts.** - Can developers add handwritten parts that are used instead of particular generated parts?

Depending on the developer's requirements it can be necessary to adapt particular parts of the generated functionality. This can be done by integrating handwritten code that refines these parts. A benefit of such handwritten code refinements is that the code generator does not have to be changed to fit different requirements.

**C3: Extendability of the generated interfaces.**
- Can the generated interfaces be extended with handwritten methods?

Hiding implementation details is accepted as common practice in software development (Parnas, 1972; Gamma et al., 1995). Accordingly, we assume that functionality of the generated system is provided to developers through dedicated generated interfaces and that the system's functionality is only accessed by using these. Obviously, these generated interfaces are oblivious to handwritten code. Consequently, the generated interfaces need to be extended to allow access to handwritten functionality.

**C4: Independence of handwritten code at generation-time.** - Is the generator independent of the existence of handwritten code at generation-time?

In some handwritten code integration mechanisms the generated code needs to be adapted if handwritten code is present. In this case the handwritten code is processed by the generator and the generated code is adapted accordingly. For instance, the handwritten code is merged into the generated code and one artifact is produced. If this functionality is not provided by the generator framework, a generator developer has to extend the generator with such functionality. This additional effort might not be desired and can be avoided by choosing a generator framework with support for handling handwritten code. However, then the choice of an integration mechanism influences the choice of the generator framework. This is not always feasible.

**C5: Independence of additional OOP language constructs.** - Can the integration mechanism be realized using only default OOP language constructs?

Some of the existing handwritten code integration mechanisms are tailored to a particular type of OOP language that provides specific language constructs. Consequently, such integration mechanisms are restricted to generators that generate code in one of these languages. The benefit of handwritten integration mechanisms using default OOP language constructs is that no additional tooling is required and the generator is not tailored to a specific type of language. The following language constructs are regarded as the default OOP language constructs in this work: (abstract) classes, inheritance, interfaces, object creation facility and message-passing capability. Except for interfaces this understanding complies with (Eliens, 1994). We are aware that not all OOP languages provide the concept of interfaces but interfaces can be realized using classes with empty method bodies and inheritance.

## 3 INTEGRATION MECHANISMS BASED ON LANGUAGE CONCEPTS

In this section, we present a catalog of integration mechanisms that presuppose certain concepts in the target language, for instance, inheritance known from object-oriented programming. Each presented mechanism is described and evaluated with the following scenario:

Assume the input model for the code generator is an UML class diagram (CD) containing the class `NotePad`. As CDs do not model class behavior, implementations of `NotePad` methods can be developed manually and the resulting handwritten code needs to be integrated with the code generated for `NotePad`.

The same problems arise whenever modeling languages do not support modeling of all aspects of a system and these parts have to be developed manually. The mechanisms presented in the remainder of this publication are not limited to CDs and a particular type of code generation. We only use CDs to give an illustrative example.

### 3.1 Generation Gap

The generation gap mechanism (Vlissides, 1998; Stahl and Völter, 2006; Fowler, 2010) assumes that an interface and a default implementation are generated for each class in the input model. For instance, the interface `NotePad` and the default implementation `NotePadBaseImpl` are generated for the class `NotePad`. Manual extensions of specific methods or behavior different from the default implementation are defined in the handwritten class `NotePadImpl`. Figure 1 depicts this pattern for the class `NotePad`. Please note, that here and in the following «gc» denotes generated code and «hc» denotes handwritten code.

In this case, `NotePadImpl` is the implementation that will be used by both the generated code as well as manually written code that uses the interface `NotePad`.

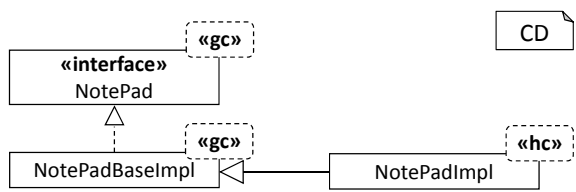

Figure 1: Generation gap pattern for the NotePad Example.

Please note, that the generation gap mechanism requires developers to create the handwritten class, no matter whether handwritten code is inserted into that class or not. In projects in which handwritten extensions are rarely needed, this leads to bloated projects with an unnecessary high number of artifacts.

**Evaluation criteria**

**C1: Separation of generated and handwritten code.** Fulfilled. This criterion holds by definition of the pattern, as the handwritten code has to be stored in separate classes.

**C2: Support for overriding generated parts.** Fulfilled. The possibility to override generated methods is a crucial feature of the pattern.

**C3: Extendability of the generated interfaces.** Unfulfilled. This approach does not provide means to reflect added methods in the generated interface. The extended generation gap mechanism (see Section 3.2) addresses this issue.

**C4: Independence of handwritten code at generation-time.** Fulfilled. Whether or what handwritten code exists does not influence the code generation in any way.

**C5: Independence of additional OOP language constructs.** Fulfilled. This mechanism does not require additional OOP language constructs.

### 3.2 Extended Generation Gap

A mechanism that addresses two disadvantages of the basic generation gap mechanism (see Section 3.1) - the inability to extend the generated interface and the necessity to create an implementation class - is the extended generation gap mechanism. Since this mechanism has been developed for our particular needs, its name is not a well-known term in MDD.

The first disadvantage is addressed by allowing to add a handwritten interface on top of the generated interface, as shown in Figure 2. As the generated interface `NotePad` extends the handwritten interface `NotePadBase`, all methods added to `NotePadBase` are also available when accessing `NotePad`. However, developers do not have to add this handwritten interface. Instead, the generator checks at generation-time whether it exists. If it does exist, the generated interface will extend the handwritten interface. Consequently, the generator needs to be executed again after adding a handwritten interface to reflect this change in the generated code.

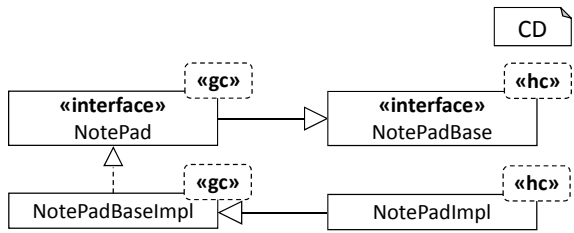

Figure 2: Extended generation gap pattern with an additional handwritten Interface.

When a developer adds a handwritten interface, the handwritten implementation class (`NotePadImpl` in Figure 2) has to be provided as well. If no handwritten interface is present, the generator generates a concrete class `NotePadBaseImpl` by default and an additional implementation class does not have to be added by developers. In this way, developers are not forced to integrate their own implementation class. However, if a developer adds the handwritten class `NotePadImpl`, which has to extend the generated base class `NotePadBaseImpl`, this class is used in the generated code and `NotePadBaseImpl` becomes abstract. This integration of handwritten code requires the generator to be executed again, as it checks at generation-time whether developers added their own implementation classes.

Other variations of the generation gap mechanism are possible. For instance, assuming that a handwritten interface always exists. A detailed discussion is neglected because the variations are very similar and, as shown in the example, differ in technical details.

**Evaluation criteria**

**C1: Separation of generated and handwritten code.** Fulfilled. See Section 3.1.

**C2: Support for overriding generated parts.** Fulfilled. See Section 3.1.

**C3: Extendability of the generated interfaces.** Fulfilled. The API of the generated class can be extended easily by adding the handwritten interface `NotePadBase` which is extended by the generated interface `NotePad`. Thus, method signatures which are added to the handwritten interface are also available in the generated one. The actual implementations of these methods have to be added to the handwritten implementation class `NotePadImpl`.

**C4: Independence of handwritten code at generation-time.** Unfulfilled. The generator has to check whether a handwritten interface or a handwritten implementation class was introduced, as this influences the structure of the generated code.

**C5: Independence of additional OOP language constructs.** Fulfilled. See Section 3.1.

### 3.3 Delegation

Delegation is a pattern of object composition in object-oriented programming. In essence, the pattern consists of two objects taking the roles of one delegator and one delegate, respectively. The delegator delegates parts of its functionality to the delegate by invoking methods of the delegate. To this end, the delegate provides an interface declaring the method signatures that can be invoked. Figure 3 gives an overview of the objects and relationships involved. Here, `NotePad` is the delegator and `NotePadDelegateImpl` is the delegate implementing the methods defined in the `NotePadDelegate` interface.

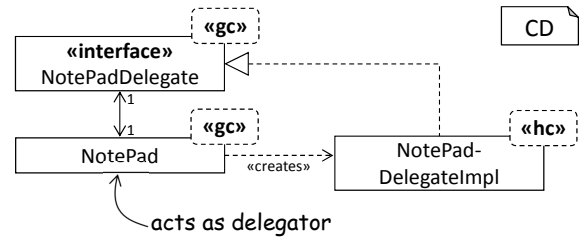

Figure 3: Delegation pattern requires a delegator for regarding handwritten implementations.

The delegator is responsible for instantiating the delegate. In this case, `NotePadDelegateImpl` is the implementation that will be used by both the generated code as well as manually written code that uses the interface `NotePadDelegate`.

The relationship between the delegator and its delegate can be unidirectional or bidirectional. In the former case, the delegate has no knowledge of its delegator. All interaction takes place via its method parameters and return values. In the latter case, the delegate has a reference to its delegator and may respond to delegated tasks by invoking methods on its delegator. In any case the delegator is in control and initiates all interactions. The delegator can choose to keep the same delegate instance for subsequent delegations or to use a new instance for every single delegation.

In essence, the purpose of delegation is to outsource functionality to a distinct object with an explicit interface specific to this functionality. This pur-pose makes delegation naturally applicable to the integration of handwritten and generated code. The roles of the delegators are taken by generated classes while the roles of the delegates are taken by handwritten classes. All functionality that cannot be generated is delegated to the handwritten delegates. The delegate interfaces, thus, are well-defined and distinct interfaces between generated and handwritten code. It can be generic and handwritten or specific and generated. The choice depends on whether the delegated functionality depends on the model or not. For instance, to delegate the implementation of method signatures in class diagrams to a delegate, it is appropriate to generate the delegate interface based on the method signatures defined by the CD.

In general, delegation provides a higher degree of encapsulation and cohesion compared to alternative patterns. Moreover, it avoids inheritance in handwritten classes since delegates only have to implement an interface. In programming languages without support for multiple inheritance delegation allows developers to use inheritance with handwritten classes.

**Evaluation criteria**

**C1: Separation of generated and handwritten code.** Fulfilled. The pattern separates generated and handwritten code by putting them into different classes and interfaces.

**C2: Support for overriding generated parts.** Unfulfilled. In this mechanism, only designated delegators can be implemented to provide handwritten code. It is not possible to override other generated parts.

**C3: Extendability of the generated interfaces.** Unfulfilled. The generated interface can be extended by subinterfaces and concrete delegators according for the extended subinterface. However, the generated delegator is not aware of these extensions.

**C4: Independence of handwritten code at generation-time.** Fulfilled. The existence of handwritten delegate classes does not influence the code generation.

**C5: Independence of additional OOP language constructs.** Fulfilled. The default OOP language constructs suffice to implement this mechanism.

**Alternatives** The cardinalities of the delegation relationship are not necessarily restricted. Thus, an implementation of the pattern may associate one delegator with exactly one delegate, or one delegator with many delegates, or many delegators with one delegate. The choice between these variants depends heavily on whether the delegate is stateful or stateless. Stateless delegates can generally be shared by many delegators and do not need to be instantiated repeatedly. Moreover, combinations of delegators and delegates can be made with different combinations of classes for both roles. For instance, a general purpose delegate may be appropriate for different classes of delegators.

### 3.4 Include Mechanism

Include mechanisms are based on dedicated language constructs which allow to define that a certain file should be included into another file at a specific point. This idea can be easily used to integrate generated and handwritten files as either a generated file includes handwritten files (see Figure 4) at designated places or vice versa. In general, the effect of using an include statement is equivalent to injecting the content of the included file to the corresponding location in the including file. Specific languages may offer include mechanisms with different meanings, but this will not be discussed in the following as the focus is on the general idea of include mechanisms.

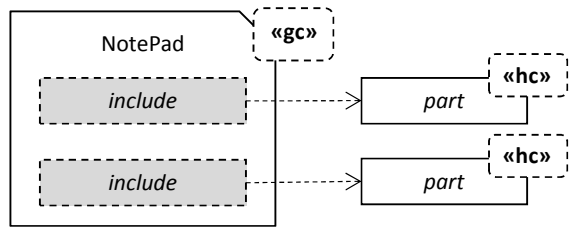

Figure 4: The include mechanism adds include statements to the generated file to consider handwritten artifacts.

By including handwritten files in generated files, the generator can define the required structure of the files and developers merely need to introduce selected handwritten files, which are included properly without the developer having to worry about it. This is advantageous if developers should not be able to deviate from this generated structure, as they can only provide the handwritten files which are included. Thus, developers are guided in which files to provide. On the other hand, if developers need more flexibility and should be able to deviate from such a generated structure, including generated files in handwritten files is more appropriate. This variant is, of course, accompanied by the risk that developers forget to include the proper generated files at the proper places.

**Evaluation criteria**

**C1: Separation of generated and handwritten code.** Fulfilled. No matter whether generated files include handwritten files or vice versa, generated and handwritten parts are separated into different files.

**C2: Support for overriding generated parts.** Unfulfilled. It is not possible to integrate handwritten code which is used instead of generated code as handwritten code can only be included and, therefore, added.

**C3: Extendability of the generated interfaces.** Conditionally fulfilled. To fulfill this criterion, a programming language has to allow include statements inside of interfaces to extend the signature. Even though we are not aware of a language that supports this, the concept itself does not forbid it. Therefore this criterion might be fulfilled, depending on the specific target language.

**C4: Independence of handwritten code at generation-time.** Fulfilled. The generation of the include functionality does not depend on the existence of handwritten code.

**C5: Independence of additional OOP language constructs.** Unfulfilled. The mechanism requires include constructs which do not belong to the default OOP language constructs.

## 3.5 Partial Classes

Partial classes facilitate splitting class implementations into several source code files. These parts are merged in a pre-compilation phase. The result contains the union of all methods, fields and super types of all its partial definitions. In contrast to aspect-oriented programming (see Section 3.6), partial classes are concerned with only one class rather than multiple.

The partial classes mechanism suits well for integrating handwritten and generated code. Each generated partial class can be extended by adding handwritten code in its own partial class in a separate source file. The resulting generated and handwritten code is integrated automatically by merging them. This merging can either be done by applying naming conventions, i.e., partial classes with the same name are merged, or by explicit notations. How and which partial classes are merged is defined by the used language.

The CD in Figure 5 illustrates the partial class mechanism. In this case, the generated code is stored in the partial class `NotePadBaseImpl` and the handwritten code is stored in a separate partial class `NotePadBaseImpl`.

**Evaluation criteria**

**C1: Separation of generated and handwritten code.** Fulfilled. The handwritten and generated par-tial classes can be located in different source code files.

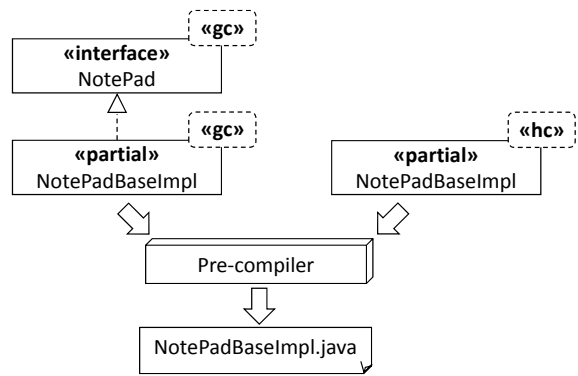

Figure 5: Partial classes mechanism merges the handwritten and the generated implementation to one single artifact.

**C2: Support for overriding generated parts.** Conditionally fulfilled. In general, the partial classes mechanism does not forbid to override methods' implementations. However, depending on the used programming language that supports partial classes, this criterion may not be fulfilled.

**C3: Extendability of the generated interfaces.** Conditionally fulfilled. The concept of the partial classes mechanism can be applied to interfaces, too. Thus, the additional method signatures can be added to the handwritten partial interface which is merged with the generated partial interface. However, if a realization of the partial class mechanism supports partial classes but not partial interfaces, this criterion is not fulfilled.

**C4: Independence of handwritten code at generation-time.** Fulfilled. Handwritten code does not have to exist at generation-time, because its existence does not influence the code generation process. Handwritten code only has to be available when the pre-compiler merges the generated and handwritten code.

**C5: Independence of additional OOP language constructs.** Unfulfilled. This mechanism requires support for partial classes which is not regarded as a default OOP language construct in this work.

## 3.6 Aspect-Oriented Programming

Aspect-oriented programming (AOP) (Kiczales et al., 1997) addresses crosscutting concerns - functionality or features scattered across several classes causing duplication - by encapsulating duplicated code in one place. Although integrating handwritten code with generated code does not necessarily deal with crosscutting concerns, AOP can be used for this integra-tion (Schindler, 2012). One advantage in this context is that the generated code does not need to offer a specific architecture to be extendable by handwritten code. Instead, the handwritten code is added by so called *aspects*.

An aspect reacts to a predefined event (*pointcut*) during the program execution. Such a predefined event can be, for instance, a method call of a specific method in a specific class. The action that is executed when a pointcut is reached is implemented in an *advice*. Such an advice can be executed before, after or instead of the according event.

The integration of handwritten code can, thus, be performed by implementing an advice that is executed instead of a specific generated method. Accordingly, handwritten code could be executed before or after executing particular generated code, by using a before or after advice. All these cases, of course, require the generator to create at least a dummy implementation of the corresponding method so that the handwritten advice can be executed instead of that generated method.

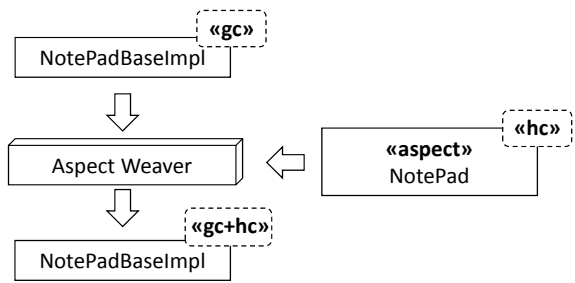

Figure 6: Overview of an aspect-oriented integration mechanism for a part of a generated software system.

Figure 6 illustrates the idea underlying the integration of handwritten code using AOP. In this case, handwritten code is added to an advice in an aspect. An aspect weaver then takes the instructions given in the aspect and produces a combined artifact (e.g. a source code file) in which the advice implementation is woven into the code of the generated classes. This means that the advice instructions are introduced into the proper locations in the generated classes. In the example given in Figure 6, the implementation of a generated method in `NotePadBaseImpl` would be replaced by the advice implementation, if the aspect contains one advice for one method.

Besides the additional overhead of weaving the aspects into the source code, a major drawback of AOP is that it is more difficult to understand the program flow as it is influenced by aspects. Moreover, refactorings in the source code may lead to invalid aspects, known as the fragile pointcut problem (Kellens et al., 2006).

**Evaluation criteria**

**C1: Separation of generated and handwritten code.** Fulfilled. AOP offers a clear separation of the handwritten and the generated code. The generated code is not aware of the aspects, which contain the handwritten code and which are stored in separate files.

**C2: Support for overriding generated parts.** Fulfilled. As described above, an advice can be implemented such that it is called instead of a particular event in the generated class. By means of this, the execution of a generated method can be prevented and instead the handwritten implementation is executed.

**C3: Extendability of the generated interfaces.** Fulfilled. Concepts in aspect-oriented programming allow to extend interfaces and classes. Consequently, the API can be extended.

**C4: Independence of handwritten code at generation-time.** Fulfilled. The generator and the generated code is not aware of handwritten code at all.

**C5: Independence of additional OOP language constructs.** Unfulfilled. In order to be able to use this mechanism, the generated code needs to conform to a programming language that supports aspect-oriented programming or contains aspect-oriented extensions. This is not provided by OOP languages by default.

**Alternatives** If the target language does not support aspect-orientation, hook points can be created in the generated code. Every hook point is called at the beginning and end of a method execution. By using inheritance, these hook points can be used to add behavior before or after the actual method execution. The behavior can be changed completely by overriding the method representing the hook point (see generation gap in Section 3.1). To some extent, this mechanism simulates aspects in AOP but it is not able to extend the API.

# 4 GENERAL INTEGRATION MECHANISMS

Besides integration mechanisms that rely on language concepts, integration approaches free of this restriction are presented in this section. In compliance with Section 3, these approaches are evaluated with respect to the criteria described in Section 2.

## 4.1 PartMerger Mechanism

A PartMerger is a component that is capable of merging multiple files of a specific type, e.g., Java files, into one file. Obviously, this idea fits well to integrate handwritten and generated parts, as these parts can be separated into different files and later be merged by the PartMerger as shown in Figure 7.

The PartMerger mechanism is a generalization of the partial classes mechanism (see Section 3.5). In contrast to partial classes, the PartMerger can also deal with non-source code artifacts. For instance, the DSL tool bench MontiCore (Grönniger et al., 2008; Krahn et al., 2010) uses this mechanism to merge generated and handwritten Eclipse plugin.xml files and Eclipse manifest files.

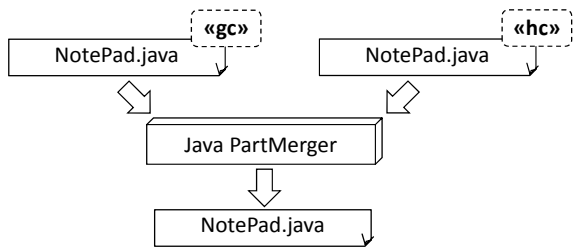

Figure 7: The PartMerger mechanism merges source code artifacts (e.g. Java source code) to one artifact.

Without any restriction on how to merge different files, the PartMerger mechanism is very flexible. To consider handwritten code, a PartMerger can give higher priorities to handwritten extensions when merging two files. Furthermore, there are different strategies for invoking the PartMerger and for defining the files to be merged. A simple strategy is to invoke the PartMerger automatically for files conforming to a specific naming convention on the artifact level, e.g., files with the same file name in specific folders or files with a common pre- or postfix. Another strategy is to let the developers configure which files should be merged.

A drawback of the PartMerger mechanism is the lack of tool support when integrating handwritten code. The reason for this is that common functionalities such as code completion are not directly available to access parts of the generated code due to the strict separation of the generated and the handwritten source code files. Developers need to implement such tooling on their own, if they want to take advantage of such tooling. This is an advantage of applying partial classes (see Section 3.5). The according tooling for this does not have to be implemented by the developer but it is already provided.

The PartMerger mechanism is very similar to the partial classes mechanism. The main difference is rooted in the language support for partial classes. In other words, a language that supports partial classes provides concepts to define what language parts are merged. The compiler takes care of the merging. In contrast, the PartMerger approach is based on a dedicated configurable tool that merges different artifacts, e.g. Java source code artifacts. Consequently, the PartMerger mechanism is not tailored to a particular language. However, the realization of the PartMerger might be tailored to particular languages.

**Evaluation criteria**

**C1: Separation of generated and handwritten code.** Fulfilled. The separation of artifacts is a precondition for this approach.

**C2: Support for overriding generated parts.** Fulfilled. A PartMerger component can be implemented in such a way that it assigns a higher priority to handwritten parts so that certain generated parts are substituted by handwritten parts in the merged artifact. In this way, the handwritten code will be executed instead of the generated code.

**C3: Extendability of the generated interfaces.** Fulfilled. Extending the API of a generated component is easily possible. To accomplish this, a handwritten interface provided by developers needs to be merged with the generated interface.

**C4: Independence of handwritten code at generation-time.** Fulfilled. Handwritten artifacts do not have to exist at generation-time, because they do not influence the generation-process. Instead, handwritten artifacts only have to be available when the PartMerger merges the generated and handwritten artifacts. This takes place after the code generator has finished.

**C5: Independence of additional OOP language constructs.** Fulfilled. This mechanism does not require any kind of OOP language construct at all.

## 4.2 Protected Regions

Protected regions are designated regions located in generated code that allow to add handwritten code (see Figure 8). A common use case for applying protected regions is to generate method signatures from input models and to insert protected regions into the corresponding method bodies.

Each protected region is typically surrounded by comments comprising a unique identification string. In this way, it can be differentiated between different protected regions. Before (re)generating code, the generator identifies protected regions in the generated code and manages the code contained in these regions based on the identification strings. While generating code, it reinserts the code previously contained in a particular protected region. As a consequence, the identification string associated with a protected region is crucial to be able to preserve the handwritten code in subsequent generator executions.

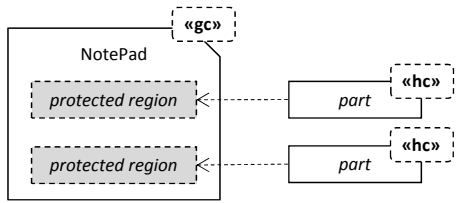

Figure 8: The protected regions mechanism requires predefined regions that contain handwritten code.

Different model-to-text transformation languages provide built-in support for declaring protected regions, including XPand (XPand website, 2014), Acceleo (Acceleo website, 2014), Epsilon Generation Language (Rose et al., 2008), JET (JET website, 2014) and MOFScript (Oldevik et al., 2005). Some of these languages have different names for the protected region mechanism, e.g., protected area (Stahl and Völter, 2006), user code block in Acceleo, user region in JET, and unprotected block in MOFScript.

A major drawback of protected regions is that it cannot be guaranteed that the generator preserves handwritten implementations. The reason for this is that handwritten code is mixed with generated code. In addition, to support this mechanism, a guarantee has to be given that the identification string is unique and stable. Otherwise, handwritten code may get lost in some situations.

**Evaluation criteria**

**C1: Separation of generated and handwritten code.** Unfulfilled. The handwritten and generated parts are mixed within the same files, therefore there is no separation according to our criterion.

**C2: Support for overriding generated parts.** Unfulfilled. It is not possible to override generated code. Only explicitly designated parts can be extended.

**C3: Extendability of the generated interfaces.** Fulfilled. An extension of the API can be achieved by generating in such a way that protected regions are introduced into the generated interfaces. Then, methods can be added to that protected region.

**C4: Independence of handwritten code at generation-time.** Unfulfilled. The generator has to analyze the previously generated code to extract handwritten code from protected regions. Otherwise the generator would not be able to inject that code from the protected regions back into the generated code.

**C5: Independence of additional OOP language constructs.** Fulfilled. This mechanism does not require any kind of OOP language construct at all.

**Alternatives** The Eclipse Modeling Framework (EMF) (Budinsky et al., 2008) applies a mechanism to integrate handwritten code which is different to protected regions described so far, but conceptually comparable. In EMF, every class, method etc. that is generated, includes a Javadoc comment that contains a generated tag (Gronback, 2009). By removing the generated tag the generated implementation can be changed. Hence, removing the generated tag corresponds to introducing a protected region.

# 5 DISCUSSION

In this section, we summarize and discuss the evaluation results shown in Table 1. A plus sign in a table cell indicates that the approach fulfills the corresponding criterion, whereas a minus sign expresses that the criterion was not satisfied. Parentheses denote that the criterion is fulfilled under certain conditions.

All approaches, except for protected regions, separate generated and handwritten code on the basis of files. Thus, it is ensured that the handwritten code is not overwritten because only the generated files are overwritten. However, the protected regions approach combines generated and handwritten code. Consequently, without external mechanisms this approach does not protect handwritten code from being overwritten.

Table 1 also shows that the extended generation gap approach, the aspect-oriented programming approach and the PartMerger approach provide the most flexibility when overriding generated parts and extending the generated interfaces. For partial classes, it depends on the actual programming language being used whether it is possible to override generated parts or not. Moreover, it also depends on the programming language, if besides partial classes also partial interfaces are supported. The situation is different for include mechanisms. Here, we are not aware of any language allowing for includes in interfaces to extend the generated interfaces. However, the concept itself does not forbid such behavior. Therefore, its applicability also depends on the used language.

A topic that should be discussed when allowing to add handwritten code is the support for restricting what parts of the generated code can be overriden. For all approaches that are based on default

Table 1: Overview of Integration Mechanisms and Results of Analysis with respect to the Criteria.

|  | Generation Gap | Extended Gen. Gap | Delegation | Include Mechanism | Partial Classes | AOP | PartMerger | Protected Regions |
|---|---|---|---|---|---|---|---|---|
| C1: Separation of generated and handwritten code | + | + | + | + | + | + | + | - |
| C2: Support for overriding generated parts | + | + | - | - | (+) | + | + | - |
| C3: Extendability of the generated interfaces | - | + | - | (+) | (+) | + | + | + |
| C4: Independence of handwritten code at generation-time | + | - | + | + | + | + | + | - |
| C5: Independence of additional OOP language constructs | + | + | + | - | - | - | + | + |

OOP language concepts restrictions can either be defined by using the private or final modifier in the generated code (generation gap and extended generation gap) or by defining a delegation interface that is to be implemented (delegation). Other approaches only allow for adding handwritten code at predefined points in the generated code (include mechanism and protected regions). For all other approaches it depends on the compiler/framework used (partial classes and AOP) or the implementation of the part merging (PartMerger) what kinds of restrictions can be defined.

All approaches but the extended generation gap mechanism and protected regions do not need to check for the existence of handwritten code at generation-time. The extended generation gap approach demands the generator to check for the existence of the handwritten interface, as the generated interface must extend this handwritten interface if it exists. Moreover, it checks for the existence of the handwritten implementation class, as this class is used in the generated code if it exists. In case of protected regions, the generator has to extract the handwritten code from the generated code before (re)generating to be able to reinsert it into the generated code.

Moreover, Table 1 illustrates that only the following approaches can be used without requiring other language constructs than the default OOP language constructs: generation gap, extended generation gap, delegation, PartMerger and protected regions. All other mechanisms depend on additional language constructs which are not provided by default by OOP languages, e.g., include functionality or partial classes.

## 6 RELATED WORK

To the best of our knowledge, no other publication exists which gives a comparable overview and evaluation of different integration mechanisms. However, as most of the presented mechanisms have been described by other authors, we give an overview of existing work in this section.

Pietrek et. al. (Pietrek et al., 2007) provide a basic introduction to MDE and list guidelines on how to integrate generated and handwritten artifacts including generation gap, protected regions, as well as include mechanisms. These mechanisms are also covered in (Petrasch and Meimberg, 2006).

Similarly, Stahl et. al. (Stahl and Völter, 2006) propose and describe the adaptation of different design patterns - in particular delegation - to integrate handwritten and generated code. These mechanisms are also discussed in (Völter, 2003; Völter and Bettin, 2004). The latter covers aspect-oriented methods as well. This concept is also employed in (Groher and Voelter, 2009; Völter and Groher, 2007; Kang et al., 2009). Additionally, a brief overview of integration mechanisms is given in (Fowler, 2010). It includes different variations of the generation gap, partial classes and protected regions mechanisms.

Approaches that specifically target .NET as the target platform and thus allow special language features including partial classes are covered in (Dollard, 2004; Greenfield et al., 2004). Both also cover protected regions, as well as generation gap and delegation.

The mechanisms supported in MetaEdit+ (Tolvanen and Kelly, 2009) are protected regions, functionality externalized in separate files (similar to the in-clude approach in our paper) as well as the direct inclusion of handwritten code in model files and are described in (Kelly and Tolvanen, 2008). Direct inclusion of handwritten code in model files has been left out, because we put the focus on integrating handwritten code in generated code.

The different approaches that affect the implementation of model-driven architecture are presented in (Frankel, 2003). They range from the unidirectional code inclusion to complete round-trip engineering, where code portions of handwritten code are traced and reflected back into the model.

The code generator LLBLGen Pro (LLBLGen Pro website, 2014) supports different target infrastructures and allows the integration of handwritten code through protected regions, user-specific templates that are included during the generation process, as well as language-specific features such as partial classes.

The approach described in (Warmer and Kleppe, 2006) supports the concept of partial classes for generated object-oriented code and protected regions for code that does not support this mechanism. Additionally, Brückmann et al. (Brückmann and Gruhn, 2010) advocate patterns such as delegation to incorporate manually written code in generated parts.

## 7 CONCLUSION

In this paper, we presented eight mechanisms to integrate generated and handwritten code for OOP languages: generation gap, extended generation gap, delegation, include mechanisms, partial classes, AOP, PartMerger and protected regions. To increase comparability between different integration mechanisms, we proposed five evaluation criteria which address different properties of integration mechanisms, for instance, whether generated and handwritten code is separated or whether it is possible to override generated parts. Table 1 on page 10 summarizes the results of the evaluation of the presented integration approaches with respect to this set of evaluation criteria.

Essentially, choosing a suitable mechanism for integrating handwritten and generated code depends on the concrete use case and the associated requirements. The catalog of integration mechanisms and evaluation criteria presented in this paper provides an overview for model-driven development tool developers that can be used to find an appropriate integration approach on a case-by-case basis. Additionally, we discussed related issues including restricting parts that can be overriden to point out concerns that have to be regarded.

We are aware of the fact that the evaluation criteria proposed in this paper might not always be sufficient to decide which integration mechanism to use. In addition, the list of presented integration approaches does not claim to be complete. However, the approaches and criteria shown in this paper can be used as a foundation that can be adapted and extended to fit specific requirements.

## REFERENCES

Acceleo website (2014). http://www.eclipse.org/acceleo/. Last visited on 22/09/2014.

Brückmann, T. and Gruhn, V. (2010). An Architectural Blueprint for Model Driven Development and Maintenance of Business Logic for Information Systems. In *Proceedings of the 4th European conference on Software architecture*, ECSA '10, pages 53–69. Springer-Verlag.

Budinsky, F., Steinberg, D., Merks, E., Ellersick, R., and Grose, T. J. (2008). *Eclipse Modeling Framework*. Addison-Wesley, 2nd edition.

Dollard, K. (2004). *Code Generation in Microsoft .NET*. Apress.

Eliens, A. (1994). *Principles of Object-Oriented Software Development*. Addison-Wesley Longman Publishing Co., Inc.

Fowler, M. (2010). *Domain Specific Languages*. Addison-Wesley.

France, R. and Rumpe, B. (2007). Model-Driven Development of Complex Software: A Research Roadmap. In *Future of Software Engineering*, ICSE '07, pages 37–54. IEEE Computer Society.

Frankel, D. S. (2003). *Model Driven Architecture: Applying MDA to Enterprise Computing*. Wiley.

Gamma, E., Helm, R., Johnson, R., and Vlissides, J. (1995). *Design Patterns: Elements of Reusable Object-Oriented Software*. Addison-Wesley Professional.

Greenfield, J., Short, K., Cook, S., and Kent, S. (2004). *Software Factories: Assembling Applications with Patterns, Models, Frameworks, and Tools*. Wiley.

Groher, I. and Voelter, M. (2009). Aspect-Oriented Model-Driven Software Product Line Engineering. In *Transactions on Aspect-Oriented Software Development VI*, pages 111–152. Springer-Verlag.

Gronback, R. C. (2009). *Eclipse Modeling Project: A Domain-Specific Language (DSL) Toolkit*. Addison-Wesley.

Grönniger, H., Hartmann, J., Krahn, H., Kriebel, S., Rothhardt, L., and Rumpe, B. (2008). Modelling Automotive Function Nets with Views for Features, Variants, and Modes. In *Proceedings of Embedded Real Time Software and Systems*, ERTS '08.

Grönniger, H., Krahn, H., Rumpe, B., Schindler, M., and Völkel, S. (2008). MontiCore: a Framework for the Development of Textual Domain Specific Languages.

In *30th International Conference on Software Engineering*, ICSE '08, pages 925–926. ACM.

JET website (2014). http://www.eclipse.org/modeling/m2t/?project=jet\#jet. Last visited on 22/09/2014.

Kang, K. C., Sugumaran, V., and Park, S. (2009). *Applied Software Product Line Engineering*. Auerbach Publications.

Kellens, A., Mens, K., Brichau, J., and Gybels, K. (2006). Managing the Evolution of Aspect-oriented Software with Model-Based Pointcuts. In *Proceedings of the 20th European Conference on Object-Oriented Programming*, ECOOP '06, pages 501–525. Springer-Verlag.

Kelly, S. and Tolvanen, J.-P. (2008). *Domain-Specific Modeling: Enabling Full Code Generation*. Wiley.

Kiczales, G., Lamping, J., Mendhekar, A., Maeda, C., Lopes, C., Loingtier, J.-M., and Irwin, J. (1997). Aspect-Oriented Programming. In *European Conference on Object-Oriented Programming*, ECOOP '97, pages 220–242. Springer Verlag.

Kleppe, A. G., Warmer, J., and Bast, W. (2003). *MDA Explained: The Model Driven Architecture: Practice and Promise*. Addison-Wesley Longman Publishing Co., Inc.

Krahn, H., Rumpe, B., and Völkel, S. (2010). MontiCore: a Framework for Compositional Development of Domain Specific Languages. *International Journal on Software Tools for Technology Transfer*, pages 353–372.

Kurpick, T., Pinkernell, C., Look, M., and Rumpe, B. (2012). Modeling Cyber-Physical Systems: Model-Driven Specification of Energy Efficient Buildings. In *Proceedings of the Modelling of the Physical World Workshop*, MOTPW '12, pages 2:1–2:6. ACM.

LLBLGen Pro website (2014). http://www.llblgen.com/. Last visited on 22/09/2014.

Navarro Pérez, A. and Rumpe, B. (2013). Modeling Cloud Architectures as Interactive Systems. In *2nd International Workshop on Model-Driven Engineering for High Performance and CLoud computing*, MDHPCL '13, pages 15–24, Miami, Florida. CEUR Workshop Proceedings.

Oldevik, J., Neple, T., Grønmo, R., Aagedal, J., and Berre, A.-J. (2005). Toward Standardised Model to Text Transformations. In *Proceedings of the First European conference on Model Driven Architecture: foundations and Applications*, ECMDA-FA '05, pages 239–253. Springer-Verlag.

Parnas, D. L. (1972). On the Criteria to be Used in Decomposing Systems into Modules. *Communications of the ACM*, 15(12):1053–1058.

Petrasch, R. and Meimberg, O. (2006). *Model-Driven Architecture: Eine praxisorientierte Einführung in die MDA*. Dpunkt Verlag.

Pietrek, G., Trompeter, J., Niehues, B., Kamann, T., Holzer, B., Kloss, M., Thoms, K., Beltran, J. C. F., and Mork, S. (2007). *Modellgetriebene Softwareentwicklung. MDA und MDSD in der Praxis*. Entwickler.Press.

Ringert, J. O., Rumpe, B., and Wortmann, A. (2013). From Software Architecture Structure and Behavior Modeling to Implementations of Cyber-Physical Systems. In *Software Engineering 2013 Workshopband*, pages 155–170. GI, Köllen Druck+Verlag GmbH, Bonn.

Rose, L. M., Paige, R. F., Kolovos, D. S., and Polack, F. A. (2008). The Epsilon Generation Language. In *Proceedings of the 4th European conference on Model Driven Architecture: Foundations and Applications*, ECMDA-FA '08, pages 1–16. Springer-Verlag.

Rumpe, B. (2011). *Modellierung mit UML*, volume 2nd Edition. Springer.

Rumpe, B. (2012). *Agile Modellierung mit UML : Codegenerierung, Testfälle, Refactoring*. Springer.

Rumpe, B., Schindler, M., Völkel, S., and Weisemöller, I. (2010). Generative Software Development. In *Proceedings of the 32nd International Conference on Software Engineering*, ICSE '10, pages 473–474. ACM.

Schindler, M. (2012). *Eine Werkzeuginfrastruktur zur agilen Entwicklung mit der UML/P*. PhD thesis, RWTH Aachen University.

Stahl, T. and Völter, M. (2006). *Model-Driven Software Development: Technology, Engineering, Management*. Wiley.

Tolvanen, J.-P. and Kelly, S. (2009). MetaEdit+: Defining and Using Integrated Domain-Specific Modeling Languages. In *Proceeding of the 24th ACM SIGPLAN conference companion on Object oriented programming systems languages and applications*, OOPSLA '09, pages 819–820. ACM.

Vlissides, J. (1998). *Pattern Hatching: Design Patterns Applied*. Addison-Wesley.

Völter, M. (2003). A Catalog of Patterns for Program Generation, Version 1.6. http://www.voelter.de/data/pub/ProgramGeneration.pdf. Last visited on 22/09/2014.

Völter, M. and Bettin, J. (2004). Patterns for Model-Driven Software-Development, Version 1.4. http://www.voelter.de/data/pub/MDDPatterns.pdf. Last visited on 22/09/2014.

Völter, M. and Groher, I. (2007). Handling Variability in Model Transformations and Generators. In *Proceedings of the 7th OOPSLA Workshop on Domain-Specific Modeling*, DSM '07. ACM.

Warmer, J. and Kleppe, A. (2006). Building a Flexible Software Factory Using Partial Domain Specific Models. In *Proceedings of the 6th OOPSLA Workshop on Domain-Specific Modeling*, DSM '06, pages 15–22. ACM.

Wile, D. S. (2003). Lessons Learned from Real DSL Experiments. In *Proceedings of the 36th Annual Hawaii International Conference on System Sciences*, HICSS '03, pages 265–290. IEEE Computer Society.

XPand website (2014). http://www.eclipse.org/modeling/m2t/?project=xpand\#xpand Last visited on 22/09/2014.